\documentclass[%
 aip,
 amsmath,amssymb,
 reprint,%
]{revtex4-1}

\usepackage{graphicx,epstopdf}
\usepackage{dcolumn}
\usepackage{bm}
\usepackage{xcolor}
\usepackage[utf8]{inputenc}
\usepackage[T1]{fontenc}
\usepackage{mathptmx}
\usepackage{float} 
\usepackage{sidecap}
\usepackage{enumerate}
\usepackage[sort&compress]{natbib}
\begin{document}

\preprint{AIP/123-QED}

\title{Soft-wall induced structure and dynamics of partially confined supercritical fluids }

\author{Kanka Ghosh}
 \email{kankaghosh@physics.iitm.ac.in}
\author{C.V.Krishnamurthy}%
 \email{cvkm@iitm.ac.in}
\affiliation{ 
Department of Physics, Indian Institute of Technology Madras, Chennai-600036, India
}%



\begin{abstract}
The interplay between structure and dynamics of partially confined Lennard Jones (LJ) fluids, deep into the supercritical phase, are studied over a wide range of densities in the context of Frenkel line (FL), which separates rigid liquidlike and non-rigid gaslike regimes in the phase diagram of the supercritical fluids. Extensive molecular dynamics simulations carried out at the two ends of the FL (P = $5000$ bar, T = $300$ K and T = $1500$ K) reveal intriguing features in supercritical fluids as a function of stiffness of the partially confining atomistic walls. The liquidlike regime of a LJ fluid (P = $5000$ bar, T = $300$ K), mimicking Argon, partially confined between walls separated by $10$ \AA{} along the z-axis, and otherwise unconstrained, reveals amorphous and liquidlike structural signatures in the radial distribution function parallel to the walls and enhanced self-diffusion as the wall stiffness is decreased. In sharp contrast, in the gas-like regime (P = $5000$ bar, T = $1500$ K), soft walls lead to increasing structural order hindering self-diffusion. Further, the correlations between structure and self-diffusion are found to be well captured by excess entropy. The rich behaviour shown by supercritical fluids under partial confinement, even with simple interatomic potentials, is found to be fairly independent of hydrophilicity and hydrophobicity. The study identifies persisting sub-diffusive features over intermediate time scales, emerging from the strong interplay between density and confinement, to dictate the evolution and stabilization of structures. It is anticipated that these results may help gain a better understanding of the behaviour of partially confined complex fluids found in nature.

\end{abstract}

\maketitle

%

Walls that are soft and flexible, play a crucial role in modifying the dynamics of the fluids and its flow behavior. The experimental evidence of the occurrence of flow instability due to fluid flow past a soft surface has been reported \cite{Kumaran2000}. Linear stability analyses have been performed for flow behavior through soft-gel coated walls \cite{dinesh2018}. Further, soft wall generated turbulence in the flow has also been studied thoroughly \cite{Srinivas2017}. These results suggest an interesting physics happening under the soft wall confinements. Furthermore, MD simulations had been carried out for water-like core-softened fluids under both fixed and flexible boundaries \cite{Krott2013}. Diffusion coefficients have been found to vary non-monotonically for core-softened fluids under fixed walls, while monotonic trend is observed for fluctuating walls \cite{Krott2013}. The structural properties of core-softened fluid are also changed while making walls fluctuating. The thermodynamic behavior of water-like anomalous fluids, modeled using core-softened potential, show a very different dependencies on the rigidity or the flexibility of the walls \cite{Bordin2014}.\\

In sufficiently thin spacings, where separation between the two confining walls are of the order of few atomic diameters of the fluid particles, interesting and anomalous behavior of fluid has been found to exist \cite{Mittal2006, Mittal2008, Vanderlick1987, Schoen1987, Schoen1988, Pagonabarraga1999, Pradeep2005, Vadhana2018, Liu2004, Magda1985}. Besides, these model systems mimic a plethora of real systems, implying a wide range of applicability from physics, chemistry and biology \cite{Carsten, Minton2001, Animesh2014}. Typically, dense fluids under confinement act quite abnormally compared to its bulk counterpart as wall-fluid interactions in a confined domain produce layer like structural patterns, which in turn significantly modify the dynamics of fluid. Even, at room temperature, extremely dilute gas under partial confinement shows appreciable changes in dynamics with a pronounced suppression of density fluctuations with respect to the bulk gaseous systems \cite{Kanka2018gas}.\\

Our recent study\cite{Kanka2018} on the effect of confinement on the structural behavior of supercritical fluids indicates the possibility of unusual dynamic behavior of supercritical fluids under confinement. Though the dynamics of supercritical fluids across the Frenkel line (FL) in bulk phase have been studied for last few years \cite{Brazhkin2012, Brazhkin2013, Prescher2017}, the domain of confined fluids in supercritical phase is vastly unexplored.\\

For the confinement, we employ a model for an atomistic wall that is aimed at capturing the thermal motion of the wall particles and ensure that the wall presents a 'rough' surface that is dynamic. The particles are arranged in a lattice as in a solid and interact with each other through a well-defined inter-atomic potential. The motion of wall atoms is coupled to a thermostat of Nose-Hoover type to maintain the same temperature as that of the supercritical fluid to avoid unnecessary heat flow through the fluid.\\
$10^5$ atoms interacting with truncated and shifted Lennard Jones (LJ) potential mimicking argon ($\epsilon$/$k_{B}$ = $120$ K, $\sigma$ = $3.4$ \AA, cutoff = $20$ \AA $\sim$ $6\sigma$) have been simulated considering a set of thermodynamics states in the supercritical regime (P = $5000$ bar,$240$ K $\leqslant$ T $\leqslant$ $1500$ K) of the bulk phase. The FL crossover has been determined through velocity autocorrelation function (VACF) and radial distribution function (RDF) calculations to be at around T $\sim$ $600$-$700$ K. We explore the changes in the dynamic and structural features across the FL through a series of MD simulations of supercritical fluid (SCF) under partial confinement using atomistic walls. To determine the FL and consistency checks in bulk, constant pressure temperature (NPT) ensembles are used while NVT ensembles are used to simulate fluids under atomistic boundaries. Walls are simulated along z, while periodic boundary conditions are applied along other two directions (x, y). Throughout the study, a fixed value of $10$ \AA{} ($\sim$ $3$ $\sigma$) has been chosen as a
\pagebreak
\onecolumngrid
\begin{widetext}
\begin{figure}[H]
\includegraphics[width=1.0\textwidth]{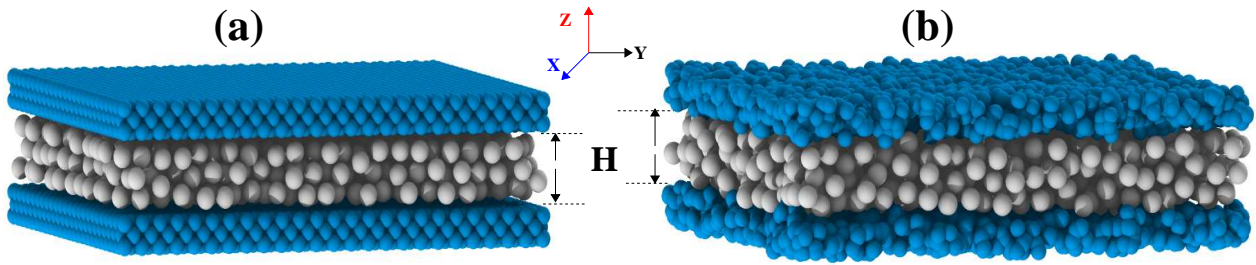}
\caption{\label{fig:diagram} A snapshot of supercritical argon confined between (a) rigid atomistic walls and (b) soft, flexible atomistic walls with a spacing H ($10$\AA) between the walls at $300$ K. Grey colored atoms denote argon and the blue atoms represent wall atoms. OVITO software is used to visualize the snapshot \cite{Ovito,Stukowski}.}
\end{figure}
\vspace{-4cm}
\end{widetext}
representative spacing to understand the interplay between the density and confinement on the dynamics and strong packing structures.
After an energy minimization, a standard velocity-verlet algorithm with a time step ($\Delta$t) of 0.0001 ps has been used to equilibrate the system up to $50$ ps followed by a $1000$ ps production run to calculate and analyze the properties of interests.\\

The solid, atomistic walls are made of three layers of the face-centered cubic (fcc) lattice. The number of wall atoms are $583444$ for T = $300$ K and $1076404$ for T = $1500$ K. Each of these wall atoms are attached to the lattice sites by harmonic springs. For our study, we chose a fixed spacing (H) of $10$ \AA ($\sim$ $3\sigma$) between the walls at both side of FL (T = $300$ K and T = $1500$ K) and vary the spring constant (k) for these springs from a higher value $5000$ ev/\AA$^2$ to a very low value of $0.005$ ev/\AA$^2$ to investigate the effect of wall softness on the structure and dynamics of supercritical fluid. Higher spring constants (k=$5000$ ev/\AA$^2$) mimic rigid walls, by restricting the MSD of the wall atoms with respect to their lattice sites to a very low value, while lower spring constants (k=$0.005$ ev/\AA$^2$) mimic soft walls, by increasing the MSD of the wall atoms with respect to their lattice sites. To define rigid wall, we allow the root mean squared displacement (RMSD) of wall particles to be $40$ times and $100$ times smaller than a typical distance traversed by a fluid particle between two collisions \cite{Brazhkin2012} ($\sim$ $1$ \AA), at $1500$ K and $300$ K respectively. Further details on parameters used for wall-wall, wall-fluid interactions, and complete computational exercises to locate FL can be found in our previous work \cite{Kanka2018}. Figure \ref{fig:diagram} shows a typical snapshot of the systems with rigid and soft walls. All MD simulations have been carried out using LAMMPS software package \cite{Steve1995, Lammps}.\\

\begin{widetext}
\begin{figure}[H]
\centering
\includegraphics[width=1.03\textwidth]{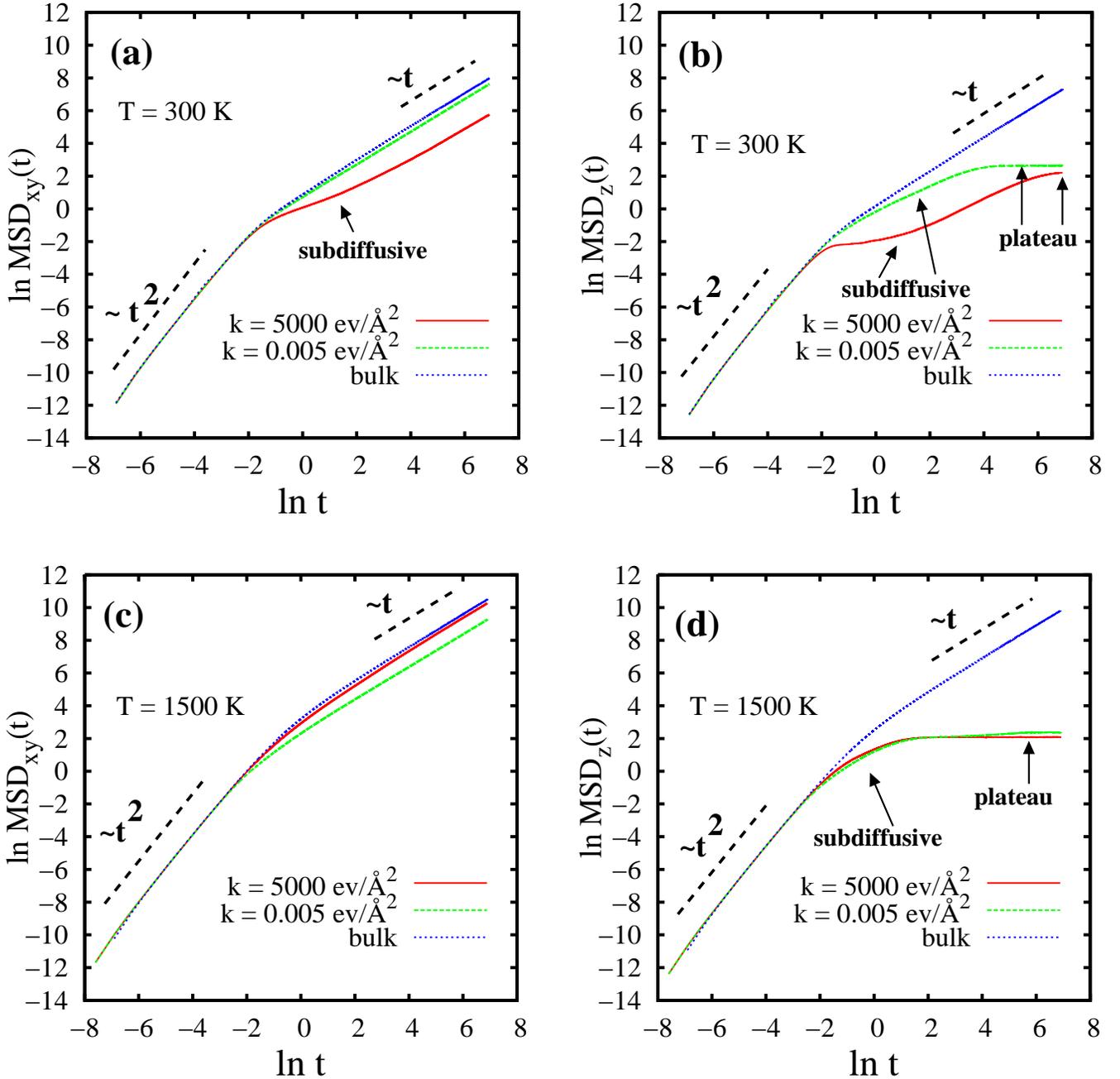}
\caption{\label{fig:2} Variation of MSD$_{\parallel}$ (along x, y) and MSD$_{\perp}$ (along z) as a function of time in a double logarithmic scale. Fig. 2.(a), (b) represent phase space point at $300$ K (liquidlike regime) for two extreme k (stiffness coefficient of walls) values and Fig. 2. (c), (d) represent phase space point at $1500$ K (gaslike regime) for same two extreme k values. For all the cases, corresponding bulk values are shown in blue dotted lines.}
\end{figure}
\end{widetext}

Figure \ref{fig:2} presents the mean squared displacement (MSD) components, parallel and perpendicular to the walls, for the liquid-like (T = $300$ K) and gas-like (T = $1500$ K) regimes of the partially confined SCF. The main results, shown for rigid (k = $5000$ eV/\AA$^2$) and soft (k = $0.005$ eV/\AA$^2$) atomistic wall conditions, feature a ballistic regime ($\sim$ t$^2$) and a diffusive ($\sim$ t) or sub-diffusive ($\sim$ t$^\alpha$; $\alpha$ = $0$) regime separated by an intermediate regime displaying a variety of cross-over characteristics. Figure \ref{fig:2} includes the bulk MSD for purposes of comparison.\\

On short time scales, the fluid particles appear to be largely unaffected by the confinement. While in the bulk, particles undergo thermal collisions resulting in diffusive motion, under partial confinement, particles undergo wall-mediated collisions in addition to thermal collisions resulting in diffusive or sub-diffusive motion. These wall-mediated collisions, being governed by the fluid density, temperature, spacing between the walls and the stiffness of the walls, are seen to produce significant changes in the MSD over intermediate and long-time scales.\\

The wall-mediated collisions are seen to constrain the z-component of the particle motion completely in the gas-like regime at T = $1500$ K (Fig.\ref{fig:2}.(d)) over long time scales. The constraint arises due to frequent velocity reversals produced by wall-mediated collisions reinforced by the strong confinement (see Velocity autocorrelation function (VACF) in Fig S1.(c) in the supplement where such velocity reversals produce minima in the VACF) that lead to progressively smaller changes in the z-component of the particle position. The effect of the wall stiffness manifests on intermediate time scales influencing the details of the cross-over from ballistic to diffusive or sub-diffusive motion. Onset of corrections to ballistic motion due to wall-mediated collisions is earlier for soft wall than for rigid wall (green and red curves in Fig \ref{fig:2}.(d)) due to the larger amplitude of thermal motion of wall particles in the former.\\

The motion parallel to the walls in the gas-like regime (T = $1500$ K) is diffusive over long time scales. Since the particles of the soft wall have a much larger root mean squared displacement than those of the rigid wall at T = $1500$ K, the wall-mediated collisions from the soft wall are significantly more non-specular in nature. Consequently, the in-plane scattering due to wall-mediated collisions induce faster decay in velocity autocorrelation function when the wall is soft than when the wall is rigid (see VACF in Fig S1.(d) in the supplement). It may indeed be noted from Fig S1.(d) of the supplement, that partial confinement induces faster decay than what is possible through purely thermal collisions. What is more interesting 
\onecolumngrid
 
is that as the wall stiffness decreases, the non-specular nature of the wall-mediated collisions increases and results in the VACF$_{\parallel}$ developing almost a shallow minimum and vanishing much earlier. The diffusion coefficients parallel to the walls in the gas-like and liquid-like regimes evaluated over a range of wall stiffness are shown in Table \ref{table1}.

As can be seen from Table \ref{table1}, at $1500$ K, the diffusion coefficients parallel to the walls (D$_{\parallel}$) are reduced by about $20$ $\%$ and $70$ $\%$ from the bulk when the wall is rigid and soft respectively. While in-plane scattering could contribute to a lowering of the diffusion coefficient, it will be seen later that an increased structural order also plays a role in the net 

\begin{table}[H]
\caption{\label{table1} Variation of Diffusion coefficients parallel to the walls (D$_\parallel$) with different wall-stiffnesses (k values) for a fixed confined spacing (H =$10$\AA) at $300$ K and $1500$ K. For reference corresponding bulk diffusion coefficients are also shown in column 4 for these two phase space points.}
\begin{ruledtabular}
\begin{tabular}{cccc}
 & k (ev/\AA$^2$) & $D_{\parallel}$(\AA$^2$/ps)  & $D_{bulk}$(\AA$^2$/ps) \\
\hline
T = 300 K &5000 &0.076 & \\
(liquidlike regime)&1000 &0.077 & \\
 &10 &0.129 &0.74 \\
H = 10 \AA & 0.5 & 0.234 & \\
& 0.05 & 0.358 & \\
& 0.005 & 0.509 & \\
\hline
T = 1500 K & 5000 & 7.191 & \\
(gaslike regime)& 1000 & 7.118 & \\
 & 10 & 6.489 & 9.117 \\
H = 10 \AA & 0.5 & 5.861 & \\
& 0.05 & 4.366 & \\
& 0.005 & 2.656 & \\
\end{tabular}
\end{ruledtabular}
\end{table}
reduction of the diffusion coefficient. \\

In the liquidlike regime (T = $300$ K), we note surprising trends in both parallel and perpendicular components of the motion particularly over intermediate time scales. From Table \ref{table1}, we note that bulk diffusion coefficient is significantly lowered with respect to that in the gas-like regime (T = $1500$ K) due to the higher fluid density. The velocity reversals, along the z-axis, produced by wall-mediated collisions acting as it does at liquid-like densities lead to caging effects over intermediate time scales and vanishes soon after for both rigid and soft walls. It is important to note that while the caging effect exists even in the bulk SCF due to the high densities, it is more pronounced when the wall is rigid than when the wall is soft (see VACF Fig S1.(a) in the supplement). This pronounced caging effect seems to arise from a nearly specular nature of wall-mediated collisions induced by the rigid wall and compounded by the strong confinement at such high densities. The surprising aspect of this pronounced caging effect is that it manifests as a sub-diffusive feature in the MSD$_{\perp}$ resembling that of a glassy phase and persists over a fairly wide intermediate time scale suggestive of the characteristic slowing down encountered in the glass forming liquids \cite{Vadhana2018}. The sub-diffusive feature exhibited by MSD$_{\perp}$ for the soft wall confinement which develops more gradually over time is distinctly different from that for the rigid wall although persisting over almost as wide an intermediate time scale. Over long time scales, however, the z-component motion is constrained for both rigid and soft wall conditions.\\

Motion parallel to the walls exhibits surprising features as well. At T = $300$ K, while the particles of the soft wall and the rigid wall would both have relatively smaller root mean squared displacements than at T = $1500$ K, the relatively smaller root mean squared displacement associated with the rigid wall would induce wall-mediated collisions that are almost specular in nature. With strong confinement compounding the nearly specular scattering, the VACF$_{\parallel}$ exhibits deeper minima as wall stiffness is increased at such high densities (see VACF Fig S1.(b) in the supplement). The existence of a well-defined minimum, followed by the vanishing of the VACF$_{\parallel}$ soon after, manifests as a sub-diffusive feature in the MSD$_{\parallel}$ when the wall is rigid. From Table \ref{table1}, we note that the diffusion coefficient is nearly $90$ $\%$ lower than that in the bulk, under rigid wall confinement, while the diffusion coefficient is only $30$ $\%$ lower than that in the bulk, under soft wall confinement. Such a large reduction of the diffusion coefficient appears possible only if in-plane scattering produced by wall-mediated collisions combines with an increased structural order imposing restrictions to particle motion.\\

The constraints on MSD$_{\perp}$ at T = $300$ K as well as at T = $1500$ K, have important structural implications – at long time, the fluid particles have root mean squared displacements of $2.9$ \AA{} and $3.7$ \AA{} at T = $300$ K, and $2.8$ \AA{} and $3.3$ \AA{} at T = $1500$ K, for the rigid and soft wall confinements respectively along z. Although the separation between the walls is $10$ \AA,
\onecolumngrid
\begin{widetext}
\vspace{-0.5cm}
\begin{figure}[H]
\centering
\includegraphics[width=0.9\textwidth]{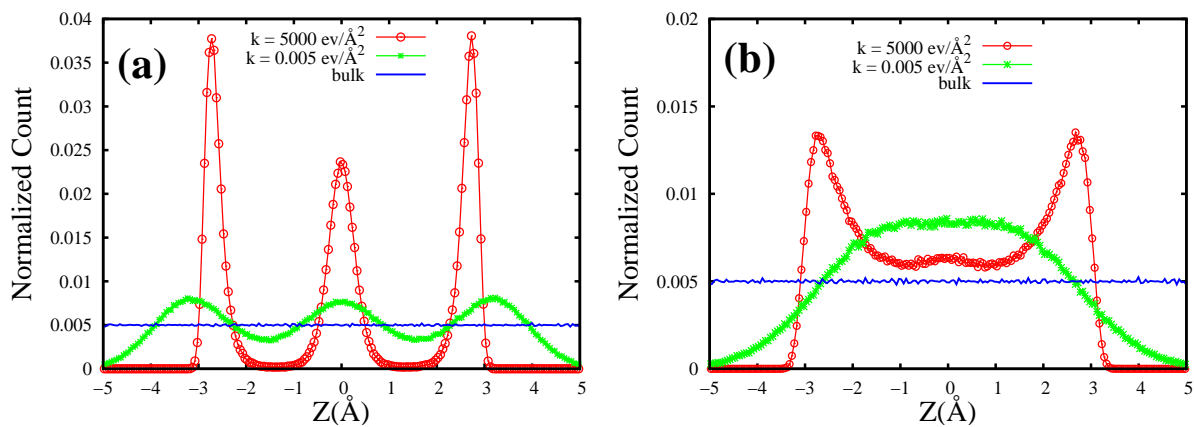}
\caption{\label{fig:3}Distributions of argon particles in supercritical state, averaged over several timesteps at $300$K (Fig \ref{fig:3}.(a)) and at $1500$ K (Fig \ref{fig:3}.(b)) are calculated for two extreme k values, normal to the walls for $10$\AA{} confined spacing. For comparison, the bulk density distribution is shown.}
\end{figure}
\vspace{-0.5cm}
\end{widetext}
these constraints on the root mean squared displacements of fluid particles force them to adopt non-uniform but symmetric spatial distributions with respect to the z-axis.  Fig \ref{fig:3}.(a) and (b) show the normalized particle counts along the z-axis at T = $300$ K and T = $1500$ K for rigid and soft wall confinements. At T = $300$ K, the higher fluid density imposes further packing restrictions resulting in a finer layering across the z-axis. The scale over which local number density varies is about $3$ \AA{} consistent with the average displacement of fluid particles in both the gas-like regime as well as the liquid-like regime of the SCF. As mentioned in our previous work \cite{Kanka2018}, the number of layers formed due to confinement scales linearly with the ratio of spacing to the atomic diameter ($\frac{H}{\sigma}$ = $10$/$3.4$ $\sim$ $3$) in the liquidlike regime of SCF.\\

It is intriguing to see how motion parallel to the walls influences the lateral structure. This structure-dynamics interrelationship is examined using the radial distribution function parallel to the walls, g$_{\parallel}$(r), and the excess entropy associated with it.\\

The radial distribution function parallel to the walls (g$_{\parallel}$(r)) is computed for specified intervals along the z-axis from 
\begin{align}
g_{\parallel}(r) \equiv & \frac{1}{\rho^2V} \sum_{i\neq j}\delta \left(r -r_{ij}\right)\Bigg[\theta\left(|z_{i}-z_{j}|+\frac{\delta z}{2}\right)\nonumber\\
& -\theta\left(|z_{i}-z_{j}|-\frac{\delta z}{2}\right)\Bigg]
\end{align}
where $V$ is the volume, $\rho$ is the density, r$_{ij}$ is the distance parallel to the walls between molecules i and j, z$_i$ is the $z$ coordinate of the molecule $i$, and $\delta(x)$ is the Dirac $\delta$ function. The Heaviside function $\theta(x)$ restricts the sum to a pair of particles located in the same slab of thickness $\delta z$. We have considered $\delta z$ to be same as the width of each layer. We use a uniform bin width and bin number of $80$ to calculate g$_{\parallel}$(r) for all the cases. Figure 4 and Figure 5 show g$_{\parallel}$(r) for soft and rigid wall confinements at the two temperatures, $1500$ K and $300$ K respectively. For purposes of comparison, the g(r) for the bulk phases are also included.\\

The two-body excess entropy, parallel to the walls, is defined as \cite{Nettleton1958, Truskett2000}
\begin{equation}
s^{(2)} = -2\pi\rho k_B \int [g_{\parallel}(r) ln g_{\parallel}(r) - g_{\parallel}(r) + 1]r^2 dr
\end{equation}
Table \ref{table2} shows the two-body excess entropy evaluated as a function of wall stiffness (k) for T = $300$ K and T = $1500$ K for the same spatial regions considered in the computation of g$_{\parallel}$(r). The bulk values of s$^{(2)}$ are negative since the reference is an ideal gas. The values for the confined systems are more negative due to the more ordered structures formed. As the density is kept constant representing a particular phase space point for both bulk and confined systems, the quantity $\frac{s^{(2)}}{2\pi\rho k_B}$ is used for the study.

\onecolumngrid
\begin{widetext}
\vspace{-0.5cm}
\begin{figure}[H]
\includegraphics[width=1.05\textwidth]{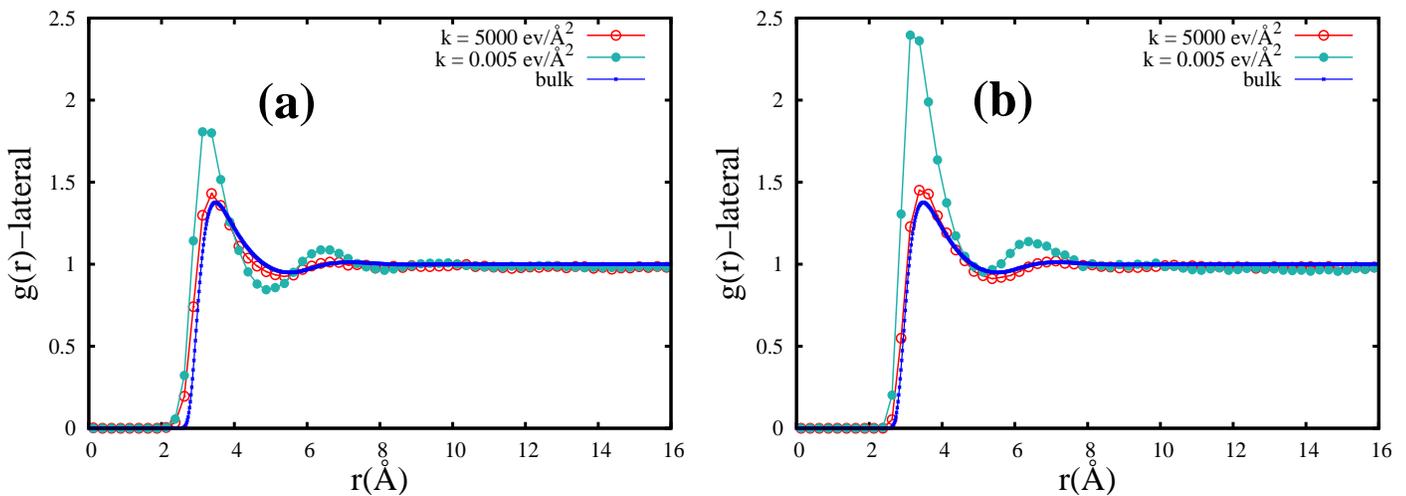}
\caption{\label{fig:4} Figure \ref{fig:4}.(a) shows g$_{\parallel}$(r) of the central region in the number distribution of SCF (argon) particles at $1500$ K, with H = $10$\AA, for two extreme stiffness coefficients (k). Figure \ref{fig:4}.(b) shows g$_{\parallel}$(r) of the region close to the left wall (wall at z = $-5$ \AA) for two extreme k values. For comparison bulk g(r) has also been shown.}
\end{figure}
\vspace{-0.5cm}
\end{widetext}
\pagebreak

Although Fig \ref{fig:4}.(a) and (b) show that g$_{\parallel}$(r) appears not to be influenced significantly by the cross-sectional density variation under rigid wall confinement at $1500$ K, compared to the bulk, the excess entropy values from Table \ref{table2} clearly indicate that there is some degree of ordering in the central region and even more ordering in the higher density region of the cross-section (see Fig. \ref{fig:3}.(b)). It may be recalled that at T = $1500$ K, under rigid wall confinement, the perpendicular component of  
\begin{figure}[H]
\centering
\includegraphics[width=0.35\textwidth,angle=-90]{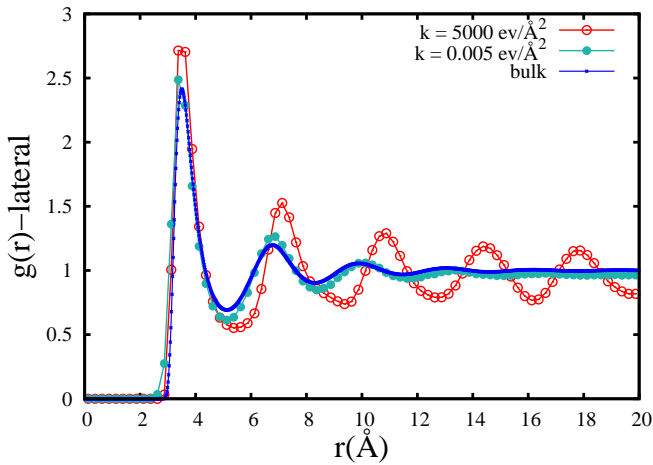}
\caption{\label{fig:5} Figure \ref{fig:5} shows g$_{\parallel}$(r) of the layer close to the left wall (wall at z = $-5$ \AA)  in the number distribution of SCF (argon) particles at $300$ K, with H = $10$\AA, for two extreme stiffness coefficients (k). For comparison bulk g(r) has also been shown.}
\end{figure}
the wall-mediated collisions is seen to (a) bring about deviations from the ballistic regime a little earlier than that in the bulk, and (b) limit the values of MSD$_{\perp}$ soon after, within a short time. This rapid change in the nature of motion is seen to produce cross-sectional density variations early on. In other words, the long-time scale density variations shown in Fig \ref{fig:3}.(b) appear to set in much earlier. The increase in structural order, captured by the lowering of the excess entropy, leads to restricted particle motion contributing to the overall reduction of $20$ $\%$ in the diffusion coefficient with respect to the bulk. The enhanced non-specular wall mediated collisions along with an increased structural order (seen in Fig \ref{fig:4} and Table \ref{table2}) parallel to the soft walls at $1500$ K, play a crucial role in the decrement of the diffusion coefficient to $70$ $\%$ compared to the bulk. \\

At T = $300$ K, the persistence of sub-diffusive features over a wide intermediate time scale seen in the MSD$_{\perp}$ appears to be the key to the formation of spatial structures and signal the mutual influence between particle motion and structure as time evolves. The `slowing down' of the dynamics implies the emergence of spatially ordered structures which in-turn restricts the dynamics and this in turn reinforces the ordering and so on till MSD$_{\perp}$ no longer changes. As can be seen from Fig \ref{fig:3}.(a), the cross-sectional density variations are stronger when the wall is rigid due to the combined effect of high fluid density and strong confinement.\\

The onset of the sub-diffusive features can be seen in the parallel component of MSD (MSD$_{\parallel}$) as well when the wall is rigid. The large reduction in the excess entropy indicates significant structural ordering and is indeed evident from g$_{\parallel}$(r) shown in Fig \ref{fig:5}. The appearance of several peaks in g$_{\parallel}$(r) and the distorted features of these successive maxima and minima suggest that the structure is more amorphous than liquid-like. The $90$ $\%$ reduction in the diffusion coefficient appears to be largely due to the higher degree of structural order.\\

When the wall is soft, the features in g$_{\parallel}$(r) appear relatively smoother with fewer peaks resembling a more liquid-like phase. Table \ref{table2} indicates that there is considerable structural ordering even when the wall is soft and appears to play a major role in reducing the diffusion coefficient by $30$ $\%$ with respect to the bulk. This again seems to be an outcome of the combined effect of high fluid density and strong confinement.\\

We note that all the results described in the article have been obtained with the ratio of fluid-wall ($\epsilon_{w-f}$) and fluid-fluid
\onecolumngrid
\begin{widetext}
\begin{table}[H]
\vspace{-0.7cm}
\caption{\label{table2} Variation of Pair-excess entropy (s$^{(2)}$/$2\pi\rho k_B$), parallel to the walls,  with different wall-stiffnesses (k values) for a fixed confined spacing (H =$10$\AA) at $300$ K and $1500$ K. For reference corresponding bulk values are also shown in column 6 for these two phase space points.}
\begin{ruledtabular}
\begin{tabular}{cccccc}
 & k (ev/\AA$^2$) & s$^{(2)}$/$2\pi\rho k_B$  & s$^{(2)}$/$2\pi\rho k_B$  & s$^{(2)}$/$2\pi\rho k_B$ & s$^{(2)}$/$2\pi\rho k_B$  \\
& &(central region) & (Region near left wall) &(Region near right wall) & (Bulk) \\
\hline
\\
T = 300 K &5000 &-63.82 &-53.48 &-52.54  \\
(liquidlike regime)&1000 &-65.31 &-53.41 &-54.31  \\
 &10 &-61.94 &-57.64 &-56.86\\
H = 10 \AA & 0.5 & -43.11 &-58.32 &-56.29 &-9.39\\
& 0.05 & -20.16 &-23.11 &-23.03\\
& 0.005 & -13.68 &-16.64 &-17.75\\
\hline
\\
T = 1500 K & 5000 & -4.86 &-7.71 &-6.70\\
(gaslike regime)& 1000 & -5.43 &-5.85 &-5.56 \\
 & 10 & -5.05 &-6.77 &-6.38\\
H = 10 \AA & 0.5 & -6.11 &-5.26 &-6.98 &-3.27\\
& 0.05 & -4.67 &-6.35 &-6.13\\
& 0.005 & -7.62 &-11.27 &-11.50 \\
\end{tabular}
\end{ruledtabular}
\end{table}
\end{widetext}
($\epsilon_{f-f}$) interaction strengths  $\frac{\epsilon_{w-f}}{\epsilon_{f-f}}$ = $1$. A value $\geq$ $1$ for this ratio represents hydrophilic boundary conditions \cite{Kalliadasis2017, Voronov2006}. Studies were undertaken explicitly varying the ratios of fluid-wall and fluid-fluid interaction strengths, mimicking both hydrophobic ($\frac{\epsilon_{w-f}}{\epsilon_{f-f}}$ $<$ $1$) and hydrophilic ($\frac{\epsilon_{w-f}}{\epsilon_{f-f}}$ $\geq$ $1$) boundary conditions. It is observed that the key findings regarding the structure and dynamics remain unaltered, thus making the results presented here quite generic and fairly independent of the affinity between the fluid and the wall particles.\\

In summary, soon after the ballistic regime, the highly dense SCF at T = $300$ K exhibits persisting in-plane sub-diffusive features under rigid-wall confinement enabling amorphous-like structural features to emerge. Over long-time scales, the structure stabilizes as evidenced by the appearance of multiple distorted peaks in the radial distribution function and the associated large negative excess entropy denoting ordering. The structure causes a nearly $90$ $\%$ reduction in the diffusion coefficient from the bulk. Perhaps not surprisingly, soft-wall confinement leads to a more liquid-like structure to emerge with only a $30$ $\%$ reduction in the diffusion coefficient from the bulk.\\ 

At T = $1500$ K, the gas-like SCF, although still dense, exhibits mild in-plane sub-diffusive features which are influenced significantly by the soft-wall confinement in sharp contrast with the behaviour at T = $300$ K. The diffusion coefficient under soft-wall confinement is reduced by nearly $70$ $\%$ from the bulk, while that for rigid-wall confinement is reduced only by $20$ $\%$ from the bulk. \\ 

The dynamics and structure of SCF under rigid and soft walls at two phase space points lying on either side of the FL, emerges as a unique system, showcasing features of almost all types of fluids: (a) liquid-like packing with prominent layering across the confined width before crossing FL, (b) gas-like diffusion dominated motion after crossing FL, (c) glass-like slowdown in dynamics at intermediate timescale and (d) amorphous-like structural features for rigid wall confinement in the liquid-like regime. \\

\section{Supplementary Material}

In the supplementary material, detailed descriptions of velocity autocorrelation function (VACF) have been discussed for supercritical LJ fluid under both rigid and soft walls at two ends of the FL (300K and 1500 K). \\

We acknowledge the help of HPCE, IIT Madras for high performance computing. KG expresses his gratitude to Department of Science and Technology, Government of India for providing INSPIRE fellowship.

\nocite{*}
\bibliography{aipsamp}

\end{document}


\preprint{AIP/123-QED}

\title{Supplementary material: Soft-wall induced structure and dynamics of partially confined supercritical fluids }

\author{Kanka Ghosh}
 \email{kankaghosh@physics.iitm.ac.in}
\author{C.V.Krishnamurthy}%
 \email{cvkm@iitm.ac.in}
\affiliation{ 
Department of Physics, Indian Institute of Technology Madras, Chennai-600036, India
}%



\maketitle

%

\section{\label{sec:level1}Velocity autocorrelation function (VACF)}
We use the notation of VACF$_{\parallel}$ and VACF$_{\perp}$ to designate the VACF of SCF along parallel ($x$, $y$) and perpendicular ($z$) directions with respect to the walls. The normalized VACF$_{\parallel}$ (Z$_{xy}$) and VACF$_{\perp}$ (Z$_{z}$) can be defined as 
\begin{equation}
Z_{xy}(t) = \frac{\left\langle \sum_{j=1}^{N} \vec{v}_{x j}(t) \vec{v}_{x j}(0) \right\rangle + \left\langle \sum_{j=1}^{N} \vec{v}_{y j}(t) \vec{v}_{y j}(0) \right\rangle }{\left\langle \sum_{j=1}^{N} \vec{v}_{x j}(0) \vec{v}_{x j}(0) \right\rangle + \left\langle \sum_{j=1}^{N} \vec{v}_{y j}(0) \vec{v}_{y j}(0) \right\rangle}
\end{equation} and
\begin{equation}
Z_{z}(t) = \frac{\left\langle \sum_{j=1}^{N} \vec{v}_{z j}(t) \vec{v}_{z j}(0) \right\rangle}{\left\langle \sum_{j=1}^{N} \vec{v}_{z j}(0) \vec{v}_{z j}(0) \right\rangle}
\end{equation}
where, $\vec v_{xj}(0)$, $\vec v_{yj}(0)$, $\vec v_{zj}(0)$ and $\vec v_{xj}(t)$, $\vec v_{yj}(t)$, $\vec v_{zj}(t)$ denote velocities of $j^{th}$ particle along $x$, $y$ and $z$ directions at initial and at some later time $t$ respectively, $N$ is the total number of particles and $\left\langle ... \right\rangle$ denotes the ensemble average.\\ 

In the gaslike regime of SCF at $1500$ K, softer walls with lower k values induce a faster decay of VACF$_{\parallel}$ compared to the rigid walls with higher k values (Fig S1.(d)). This feature suggests an increasing number of non-specular collisions with soft walls (higher RMSD of wall particles) giving rise to more in-plane scattering. Comparatively, rigid walls induce slower decay in VACF$_{\parallel}$ due to lower RMSD of wall particles. VACF$_{\perp}$ exhibits minima, as the rigidity of the walls is gradually increased, due to frequent velocity reversals produced by wall-mediated collisions (Fig S1.(c)). These velocity reversals due to wall-mediated collisions, occurring in the intermediate timescales, influence the cross-over from ballistic to diffusive or sub-diffusive motion as shown in the main text. Fig S1.(d) shows that partial confinement affects VACF$_{\parallel}$ in SCF, unlike in a confined dilute gas \cite{Kanka2018gas}, and induces faster decay than what is possible through purely thermal collisions (bulk VACF).\\

Nearly specular wall-mediated collisions under rigid wall confinement combined with high density gives rise to strong caging effect. This caging effect, at $300$ K (liquidlike regime of SCF), has been found to manifest as minima in VACF$_{\perp}$, even in the bulk SCF due to the high densities, though, it is more prominent when the wall is rigid (Fig S1.(a)). This caging effect triggers a sub-diffusive feature in the MSD$_{\perp}$, as shown in the main text, resembling that of a glass-like behavior \cite{Vadhana2018} that persists over a fairly wide intermediate time scale. Both density and confinement influence VACF$_{\parallel}$ as well and minima are shown to be deeper for rigid walls (Fig S1.(b)). The caging and slowing down of dynamics in the liquidlike regime of SCF indicate emergence of structural features in g$_{\parallel}$(r) under confinement (discussed in the main text). \\
\makeatletter 
\renewcommand{\thefigure}{S\@arabic\c@figure}
\makeatother
\setcounter{figure}{0}
\begin{figure}[H]
\centering
\includegraphics[width=0.5\textwidth]{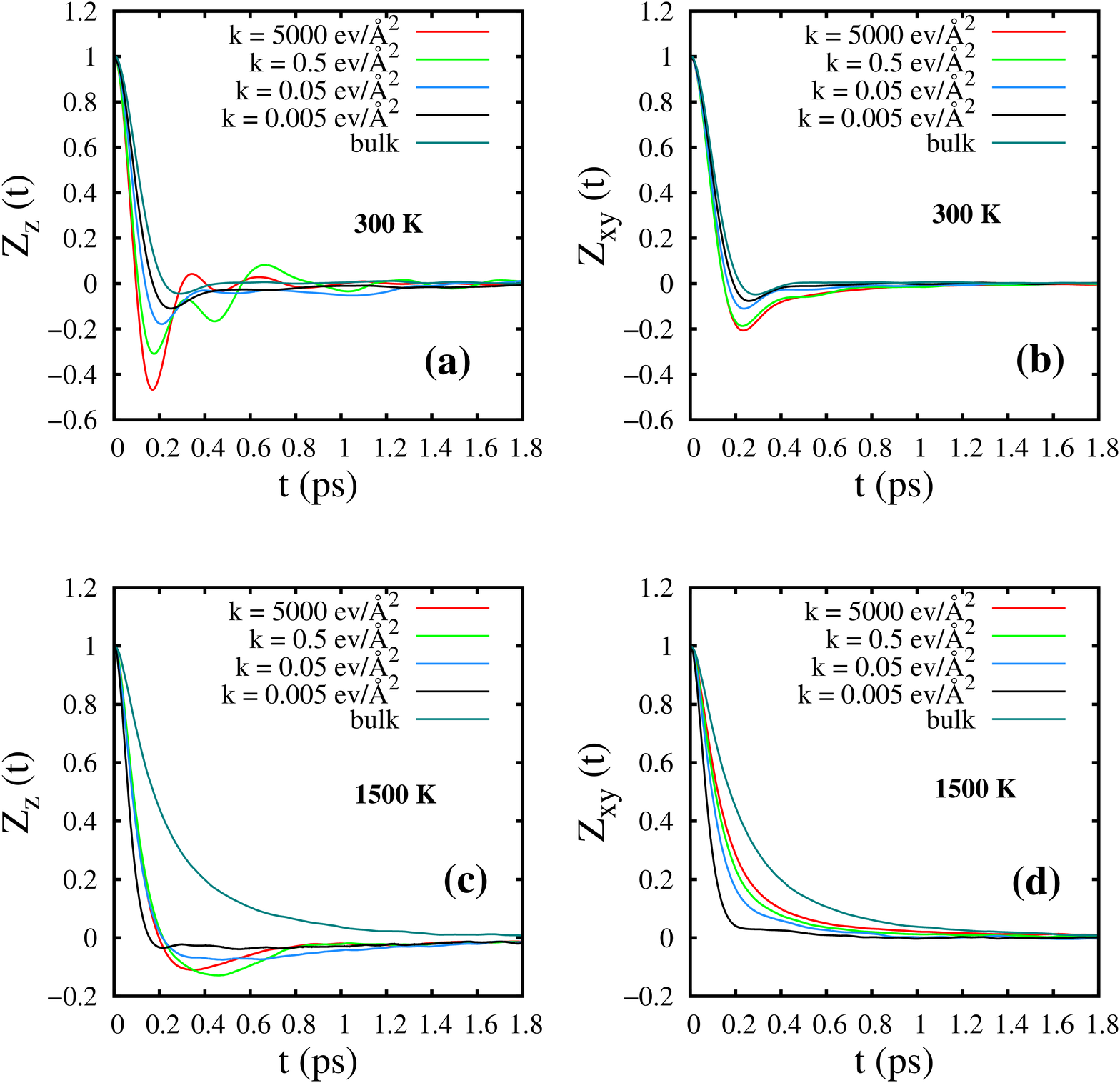}
\caption{\label{fig:s1}Normalized VACF$_\parallel$ (Z$_{xy}$) and VACF$_\perp$ (Z$_{z}$) as a function of time for different stiffness coefficients (k) of the wall atoms for a fixed H (=$10$\AA). Fig S1.(a), (b) represent phase space point at $300$ K (liquidlike regime) and Fig S1.(c), (d) represent phase space point at $1500$ K (gaslike regime). For comparison the corresponding bulk VACF are also shown in each of the cases.}
\end{figure}


\nocite{*}
\bibliography{supplement}